# A Systematic Literature Review on Continuous Integration and Deployment (CI/CD) for Secure Cloud Computing

Sabbir M. Saleh[1][a], Nazim Madhavji[1][b] and John Steinbacher[2][c]
[1]*Department of Computer Science, University of Western Ontario, London, Ontario, Canada*
[2]*IBM Canada Lab, Markham, Ontario, Canada*
ssaleh47@uwo.ca, madhavji@gmail.com, jstein@ca.ibm.com

Abstract: As cloud environments become widespread, cybersecurity has emerged as a top priority across areas such as networks, communication, data privacy, response times, and availability. Various sectors, including industries, healthcare, and government, have recently faced cyberattacks targeting their computing systems. Ensuring secure app deployment in cloud environments requires substantial effort. With the growing interest in cloud security, conducting a systematic literature review (SLR) is critical to identifying research gaps. Continuous Software Engineering, which includes continuous integration (CI), delivery (CDE), and deployment (CD), is essential for software development and deployment. In our SLR, we reviewed 66 papers, summarising tools, approaches, and challenges related to the security of CI/CD in the cloud. We addressed key aspects of cloud security and CI/CD and reported on tools such as Harbor, SonarQube, and GitHub Actions. Challenges such as image manipulation, unauthorised access, and weak authentication were highlighted. The review also uncovered research gaps in how tools and practices address these security issues in CI/CD pipelines, revealing a need for further study to improve cloud-based security solutions.

## 1 INTRODUCTION

Cloud computing has become the go-to method for software deployment because it offers clear advantages over traditional setups. These include flexible infrastructure, accessible data storage and sharing, less administrative hassle, and access from anywhere. Continuous Integration (CI), originating from Extreme Programming (XP) (Newkirk, 2002), is an Agile method where team members regularly integrate code changes, which results in faster production, better product quality, and a more effective team overall (Fitzgerald and Stol, 2017).

Automation plays a crucial role in CI, especially in testing and development. It boosts efficiency, improves teamwork among developers, and leads to more predictable releases (Leppänen et al., 2015; Ståhl and Bosch, 2014; Fitzgerald and Stol, 2014). CI, along with Continuous Delivery (CDE) and Continuous Deployment (CD), are core parts of DevOps (Lacoste, 2009). CD is about deploying software to an environment cloud, while CDE takes it further by managing updates (Humble and Farley, 2010). Automating these processes makes the process more efficient and improves software quality (Weber et al., 2016) while reducing risks (Bar et al., 2013).

While automation helps in many ways, it also brings certain security risks. Vulnerabilities such as Regular Expression Denial of Service (ReDoS) (Saboor et al., 2022) can open cloud services to attacks such as Log4j, SolarWinds, and CodeCov.

Of the 573 articles we reviewed, 66 met our selection criteria (see Section 3.3). These articles helped us explore the following research questions:

*RQ1.* What tools and methods are available for securely implementing CI/CD in the cloud?

*RQ2.* What solutions have been suggested for maintaining secure CI/CD pipelines in cloud environments?

*RQ3.* What are the main challenges when securing cloud-based CI/CD pipelines?

---

[a]  https://orcid.org/0000-0001-9944-2615
[b]  https://orcid.org/0009-0006-5207-3203
[c]  https://orcid.org/0009-0001-6572-6326

This study reviewed the current tools (Section 4.1), proposed solutions (Section 4.2), and challenges (Section 4.3) regarding secure CI/CD pipelines over the cloud platform.

To identify the challenges (Section 4.3) that prevent practitioners from adopting solutions, leading to security vulnerabilities.

The rest of the paper is structured as follows: Section 2 looks at related work, including review method and the possible research gaps (Sections 2.1 and 2.2) identified from our RQ findings. Section 3 explains the SLR method, covering RQs (Section 3.1), search strategy (Section 3.2), data sources (Section 3.3), inclusion/exclusion criteria (Section 3.4), and the SLR steps (Figure 2), along with how we extracted and synthesised the data (Section 3.5).

Section 4 presents the results, demographic data (Figure 3), and findings for each RQ (Sections 4.1, 4.2, 4.3). We follow this with analysis and discussions in Section 5. Threats to validity are covered in Section 6, and Section 7 wraps things up with conclusions and future work.

## 2 RELATED WORKS

During our SLR, we identified literature reviews, survey papers, and systematic literature reviews. These addressed various aspects of CI/CD.

Shahin et al. (2017b) surveyed CI/CD and DevOps practitioners, highlighting deficiencies in automated testing, rigid deployment methods, and security awareness. They aimed to categorise elements influencing CD practice adoption, such as better tools and management support.

Zhang et al. (2018) detailed practitioners' struggles with containerising CD and identifying prerequisites and challenges before establishing CI-based Workflow (CIW) and Docker Hub auto-builds Workflow (DHW). They noted trade-offs in stability and simplicity and the need for better security and access controls. An IDE model for cloud-based Static Application Security Testing (SAST) tools was implemented but did not significantly enhance fixing insecure code.

Waseem et al. (2021, 2023) discussed the security vulnerabilities in microservices developed with Docker that are open to cyberattacks and highlighted the need to focus on pipeline security over the cloud.

Zampetti et al. (2023) emphasised that combining hardware and software expertise can overcome CI and CDE implementation challenges in Cyber-Physical Systems (CPS), focusing on SW and HW component interactions.

Shahin et al. (2021) analysed DevOps forums to identify architecture design issues, noting that deployment, security, and testing were the most challenging during DevOps transitions.

Faustino et al. (2022) reviewed DevOps scenarios, noting faster delivery and increased automation. However, security issues have yet to be discussed.

Rajapakse et al. (2022) identified challenges and solutions for adopting DevSecOps, focusing on collaboration, insider threats, and limitations of SAST and Dynamic Application Security Testing (DAST) tools. They aimed to understand the difficulties in adopting DevSecOps.

Shahin et al. (2019) proposed a framework to re-architect CD with goals for Operational Aspects (e.g., development settings, stakeholders' requirements) and Quality Attributes (e.g., resilience, modifiability, deployability, etc.).

Shahin et al. (2017a) discussed issues in adopting CI/CD/CDE, such as coordination, skills, and tools. They also noted a need for more research on pipeline security and stability.

Table 1 presents the area between our SLR and the existing work.

Table 1: Summarising the Focused Area.

| Publications | Focused Areas |
|---|---|
| Shahin et al. 2017b | Automation of CD |
| Zhang et al. 2018 | IDE for SAST |
| Waseem et al. 2021, 2023 | Microservice |
| Zampetti et al. 2023 | Collaboration of SW and HW |
| Shahin et al. 2021 | Architectural issues in DevOps |
| Faustino et al. 2022 | Benefits of DevOps |
| Rajapakse et al. 2022 | Adoption of DevSecOps |
| Shahin et al. 2019 | Architectural issues in CD |
| Shahin et al. 2017a | Adoption of CI/CD/CDE |
| This SLR | Security of CI/CD over the Cloud |

### 2.1 Review Methodology

In software engineering (SE), conducting multiple reviews on a single topic is common (Shahin et al., 2017a). Since the introduction of Evidence-Based Software Engineering (Kitchenham et al., 2004, 2006, 2022a), systematic literature reviews (SLRs) have become a key research method (Zhang et al., 2011). However, reviewing secure CI/CD in the cloud requires a more focused approach (Düllmann et al., 2018).

### 2.2 Research Gaps

There is a growing need for research to improve security in containerised applications. This includes refining tools such as seccomp profiles for Docker,

AppArmor, SELinux, and content trust (Garg and Stavik, 2019; Le et al., 2023; Lopes et al., 2020).

Low-code platforms present security challenges, mainly due to weak authentication and cybercrime (Rafi et al., 2022).

GitHub Actions has security concerns that require further study (Decan et al., 2022; Koishybayev et al., 2022; Hilton et al., 2017; Benedetti et al., 2022a). Research into architectural challenges, such as deployment, security, and testing, is also important.

Principles like shift-left security, compliance with standards (OWASP, NIST), and zero-trust architecture can make systems more resilient (Shahin et al., 2017; Zhang et al., 2018; Shahin et al., 2021).

Finally, there is potential for new automated Software Supply Chain (SSC) solutions to detect vulnerabilities and enhance the security of CI/CD pipelines (Enck and Williams, 2022; Byrne et al., 2020; Karl et al., 2022).

# 3 RESEARCH METHOD

We conducted an SLR, which combines available research relevant to a focused area of interest and specific RQs. By following the guidelines of Kitchenham, B. et al., 2022a, our research method consists of planning, conducting, and reporting with the specification of the RQs, identifying research by generating a search strategy, selecting primary studies through inclusion and exclusion criteria, and data extraction and synthesis.

## 3.1 Goal, Question, Metric (GQM)

The Goal of this SLR is to analyse and synthesise tools and approaches for securing CI/CD pipelines on cloud services, highlight the challenges of existing solutions, and answer the RQs.

We prepared our RQs according to the criteria of the PICOC by Mark, and Helen (2008) – Population (a deployment area, e.g., the cloud), Intervention (technologies to perform specific tasks, e.g., tools), Comparison (with which the intervention is being compared, e.g., the practitioners), Outcomes (findings, e.g., existing approaches, the challenges, and the practices to the goal {secure the CI/CD pipeline over the cloud}), and Context (in which the analogy will take place, e.g., the industry).

The identified Metrics for this SLR are:

Identifying existing and proposed methods, technologies, and practices for secure CI/CD maintenance. Classifying and enumerating security challenges (e.g., gaps, integration, performance, etc.) in maintaining CI/CD pipelines in the cloud.

## 3.2 Search Strategy

Specific search phrases were created to find relevant studies based on the guidelines from Zhang et al. (2011) and Kitchenham et al. (2022a). This task faced challenges because many papers used synonyms like "cloud security" and "cybersecurity." To enhance our search, we employed snowballing (Wohlin, 2014) by examining citations in the studies and conducted a manual search as recommended by Zhang et al. (2011). This established a Quasi-Gold Standard (QGS), identifying 91 relevant papers. The initial search string was:

```
TITLE-ABS-KEY (("cloud" OR "cloud computing" OR
"cloud computing services" OR "infrastructure") AND
("security" OR "cloud security" OR "cloud computing
security" OR "cybersecurity" OR "cyber security" OR
"cyber-attack" OR "threat") AND ("continuous integration"
OR "CI") AND ("continuous deployment" OR "CD"))
```

Figure 1: Search String of the initial search for SLR.

## 3.3 Data Collection Sources

The automatic search was carried out across six digital libraries: Scopus, ACM, IEEE Xplore, Wiley, Springer Link (SL), and ScienceDirect (SD) (Chen et al., 2010).

CiteSeerX and AIS eLibrary have complex search functions and lack post-query refinements (Li & Rainer, 2022; Brereton et al., 2007). Kluwer has merged with and is indexed by Springer Link (Gusenbauer and Haddaway, 2020; Maplesden et al., 2015). Additionally, Inspec overlaps with Scopus (Maplesden et al., 2015). In contrast, Google Scholar yields results with less than 1% accuracy for systematic searches (Gusenbauer and Haddaway, 2020; Chen et al., 2010; Boeker et al., 2013).

## 3.4 Study Selection Criteria

We established inclusion and exclusion criteria to identify studies relevant to our research questions, considering these criteria might be adjusted as we moved through the search process (Staples and Niazi, 2007).

Inclusion Criteria:
- Full-text (Brereton et al., 2007) peer-reviewed papers published in English.
- Address CI/CD security in the cloud.
- Empirical research (Kitchenham et al., 2022b).

Exclusion Criteria:
- Abstracts, conference info, news, and videos.

- Earlier versions of papers by the same authors when more recent versions are available (e.g., conference vs. journal publications).
- Duplicate studies from digital libraries.

### 3.5 Data Extraction and Synthesis

We read the full text of the selected papers for review and reporting, applying the inclusion and exclusion criteria.

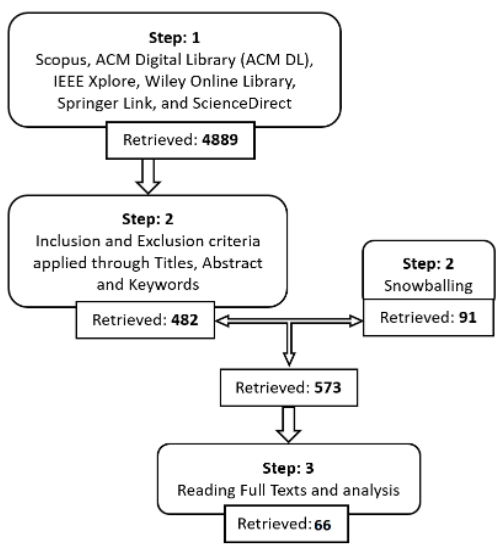

Figure 2: Steps of the Study Selection for SLR.

We passed the subsequent steps for this SLR:

*Step 1:* We started with 4,889 articles based on the search criteria.

*Step 2:* We screened the titles, keywords, and abstracts to narrow it down to 573 papers. Of these, 482 directly met our criteria, and an additional 91 were found using the Snowballing method.

*Step 3:* We reviewed the introductions and conclusions of the 573 papers, selecting those relevant to our study. After thoroughly reviewing the full articles, 66 were included in our final selection.

## 4 RESULTS

This section summarises the research questions' findings (sections 4.1, 4.2, 4.3) by synthesising and analysing the extracted data. Figure 3 displays the publication demographics, showing that from 2021 to 2023, 40 of the 66 relevant papers (over 60%) were published, emphasising the recent focus on CI/CD security in the cloud. Most of these publications appeared in conferences, with 41 papers (62.12%), followed by 15 journal articles (22.73%) and 10 workshop papers (15.15%).

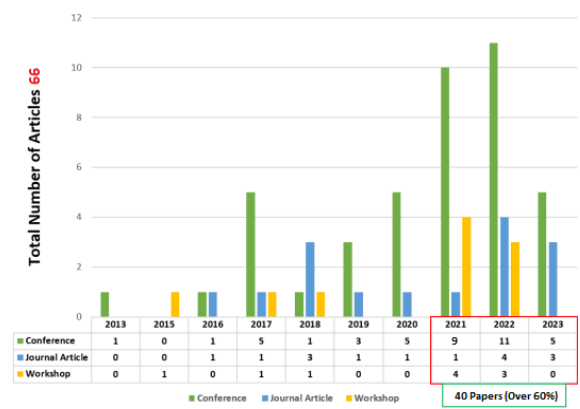

Figure 3: Demographic Data of Relevant Studies.

### 4.1 Findings of RQ1

We present the tools, approaches, and frameworks identified in our review with short descriptions (Table 2). We compiled information on 62 tools and eight distinct approaches and frameworks.

### 4.2 Findings of RQ2

Here, we list the proposed tools and approaches with short descriptions (Table 3) retrieved from the papers. We compiled information on five tools and twelve approaches/frameworks.

Some recommended practices (findings of RQ2) for organisations to address CI/CD pipeline security issues:

Trust developers: If they can make deployment decisions, it may facilitate the continuous deployment process (Shahin et al., 2017b).

Increase collaboration between operations and development teams: This may help complete complex tasks effectively (Shahin et al., 2021).

Invest in automated testing and quality assurance for continuous delivery (Shahin et al., 2017b).

Securing a software supply chain requires transparency, validity, and separation between activities and components (Okafor et al., 2022).

Providing access to developers from tool builders of Jenkins, CircleCI, TravisCI, etc., helps to provide better feedback (Hilton et al., 2017).

Limiting the CI/CD access may protect the pipeline from tampering (Pecka et al., 2022).

A solid engineering culture can emphasise quality where employees can become experts (Dursun, 2023).

Table 2: Existing Methods (approaches and frameworks) and Tools.

| Name | Description | Reference |
|---|---|---|
| Docker Bench for Security | Tool for enforcing security best practices for Docker images/containers. | Garg and Stavik, 2019 |
| Docker Trusted Registry | Secure storage and deployment of Docker images/containers. | |
| CodeShip | SaaS for logging CD workflow failures. | Zhang et al., 2018 |
| CoreOSs Clair, OpenSCAP, Anchore Engine, Trivy | Vulnerability scanners using NVD and CVEs data. | Garg and Stavik, 2019; Brandy et al., 2020; Mahboob and Coffman, 2021; Throner et al., 2021; Nadgowda and Luan, 2021 |
| SonarQube, SonarCloud | Tools for detecting security issues and maintaining code quality in CI/CD. | Abhishek and Rao, 2021; Athamnah M. et al., 2021; Luo L. et al., 2021; Romero E. at al., 2022; Leite et al., 2019 |
| Snyk | Scans dependencies to ensure trust in the Software Supply Chain (SSC) within CI/CD. | Throner et al., 2021; Bass et al., 2015; Alfadel et al., 2023 |
| CodeQL (Code Analysis Platform) | An automation tool for identifying security vulnerabilities | Alfadel et al., 2023, Okafor, C et al., 2022, Pan, Z. et al., 2023 |
| Super-Linter | A repository with multiple linter tools | Cankar et al., 2023, Chhillar and Sharma, 2019 |
| Mega-Linter | Tool to analyse CI/CD consistency | Cankar et al., 2023 |
| Prisma Compute (Twistlock), Prisma Cloud, Aqua | Container security tools for vulnerability scanning, runtime protection, and blocking unsafe builds. | Athamnah M. et al., 2021, Le et al., 2023 |
| Analizo, Code Climate | Source code analysers are used to identify vulnerabilities and bug risks. | |
| Prometheus, Zabbix, Nagios | Incident management and monitoring tools. | Leite et al., 2019 |
| Graylog, Logstash | Log management tools for security and reliability. | |
| Splunk, DynaTrace, Dapper, AppDynamics | Monitoring tools for detecting and blocking security threats. | Bennett and Barrett, 2018 |
| Veracode, LGTM, Checkmarx, CodeGuru Reviewer, FindBugs, CheckStyle, ESLint, Coverlay, IntelliJ, Coverity Scan | SAST tools to detect vulnerabilities early in SDLC. | Luo et al., 2021 |
| IBM UrbanCode Deploy, Microsoft Visual Studio Release Management | ARA (Application Release Automation) tools for identifying bugs, memory leaks, and code smells. | Révész and Pataki (2017, 2019) |
| Debricked, NSP, Sonatype, vuln-regex-detector | CI tools for scanning commits/PRs and automating vulnerability detection. | Alfadel et al., 2023 |
| Cijitter, CijScan | CI tools for defending against cryptojacking. | Alfadel et al., 2023, Li Z et al., 2022 |
| AppArmor, SELinux | Docker security tools for defence layers. | Garg and Stavik, 2019, Le et al., 2023, Lopes et al., 2020 |
| Seccomp | Restricts app access to ensure security. | Le et al., 2023, Lopes et al., 2020 |
| Spire, Dependabot, tekton-chain, Code Risk Analyzer, Mend | DevSecOps solutions play a critical role in CI/CD security. | Nadgowda and Luan, 2021 |
| Chef | OSS is used to configure and secure DevOps in the cloud. | Alonso et al., 2022 |
| ART | Autonomous real-time testing for CI/CD (DevTstOps). | Fehlmann and Kranich (2021) |
| Asylo | Development framework ensuring privacy through TEEs. | Mahboob and Coffman, 2021 |
| STRIDE | Microsoft's threat modelling framework (Spoofing, Tampering, etc.). | Davis et al., 2022 |
| Signature-based, Anomaly-based | Approaches for monitoring containers and securing CI/CD pipelines. | Jyothsna et al., 2011; Kumar and Sangwan, 2012 |
| Harbor | Blocks deployment of unscanned Docker images. | Mahboob and Coffman, 2021, Throner et al., 2021 |
| VirusTotal | Scans Docker images for malicious content. | Abhishek and Rao (2021) |
| GitHub Actions (GHA) | Automates CI/CD and mitigates security risks. | Okafor, C et al., 2022, Tu et al., 2021 |

Table 3: Proposed Methods and Tools.

| Name | Description | Reference |
|---|---|---|
| ACT Testbot Automated Continuous Testing | Automated bot for continuous testing, defect analysis, reporting, and management in CI/CD builds. | Chhillar and Sharma, 2019 |
| UBCIS | Benchmarks vulnerabilities in container scanning tools (e.g., Debian, Ubuntu, Alpine). | Berkovich et al., 2020 |
| GHAST, GWChecker | Scans GitHub Actions workflows for security weaknesses, auto-notifies for protection against SSC attacks. | Koishybayev et al. 2022, Benedetti et al., 2022a |
| CIAnalyser | Removes malicious code from OSS CI/CD scripts/pipelines. | Pan, Z. et al., 2023 |
| Multi-layered security | Framework for preventing Docker image vulnerabilities, with scanners at each pipeline layer. | Brandy et al., 2020 |
| DIVA Docker Image Vulnerability Analysis | Detects and evaluates security issues in Docker images. | Shu et al., 2017 |
| CloudInspector | Provides real-time, auditable security information in a CI/CD pipeline. | Flittner et al., 2016 |
| Cluster-Scoped-CICD | Kubernetes CI/CD pipeline with privacy guarantees using Asylo. | Mahboob and Coffman, 2021 |
| ADOC | Automated DevSecOps framework for addressing security risks with a defense-in-depth strategy. | Kumar and Goyal, 2020 |
| DVE (Deliberated Vulnerable Environment) | Stores and auto-processes exploited scripts and vulnerability data for cloud-native applications. | Huang et al., 2020 |
| Buildwatch | Monitors pipeline dependencies to detect security risks. | Ohm et al., 2020 |
| SUNSET | Identifies and evaluates software supply chain security risks. | Benedetti et al., 2022b |
| SySched | A call-aware container scheduler secures CI/CD by blocking unsafe builds and scanning for known CVEs. | Le et al., 2023 |
| Tapiserí | Visionary DevSecOps design for certification and introspection of a pipeline. | Nadgowda and Luan, 2021 |
| Blockchain Technology | Enhances pipeline security, transparency, traceability, and tamper-proofing through blockchain. | Akbar et al., 2022, Bankar and Shah 2020 |
| Supervised Learning | Machine learning is used to automate tests in CI/CD to mitigate attacks. | Drees et al., 2021 |

## 4.3 Findings of RQ 3

Below, we report the challenges in existing tools and approaches, including practices that raise security issues within cloud-based CI/CD pipelines.

**Authorisation.** Trusted Execution Environments (TEE) can enhance security, but Dev resources may be at risk if hackers can access Harbor (Mahboob and Coffman, 2021). Inadequate authorisation can result in pipeline security issues (Throner et al., 2021).

**Vulnerabilities Assessment.** This happens pre-deployment, leaving post-deployment updates unchecked and insecure (Huang et al., 2020). Due to the complexities of Infrastructure IaC, inspecting workflows for security flaws is challenging (Cankar et al., 2023; Alonso et al., 2022).

**Tools Integration.** Tools such as Clair, SonarQube, GoKart, etc. should be rapidly integrated into cloud platforms, though they require long-term commitments (Garg and Stavik, 2019; Abhishek & Rao, 2021; Christakis et al., 2022). The disconnection of tools such as Coverity Scan, LGTM, and Checkmarx from IDEs can render scanning results obsolete if the code is updated during the scan (Luo et al., 2021).

**Third-Party and OSS Tools.** Choosing consistent tools is crucial due to vulnerabilities in third-party software and OSS (Kumar and Goyal, 2020; Berkovich et al., 2020). Integrating these tools faces challenges with security boundaries, upgrade complexities, and practitioner reluctance to update, leading to outdated dependencies and security issues such as lack of authentication (Zampetti et al., 2023, Pan et al., 2023; Zhu et al., 2023; Benedetti et al., 2022b).

**Layer of Defence.** Regular updates are essential (e.g., for Seccomp) to prevent DoS attacks, but determining necessary updates is complex and time-consuming, hindering practitioner approval (Lopes et al., 2020).

**Architectural Design Issues.** Deployment, security, and testing are challenging (Shahin et al., 2017). Developers and customers have concerns about existing tools and need help with cloud deployment (Shahin et al., 2021). To address developers' pain points, better testing support and automatic security upgrades in CD workflows are required (Zhang et al., 2018).

**GitHub Actions (GHA).** While GHA can potentially reduce CI/CD pipeline security issues by recommending specific commits, it faces low adoption and has security concerns such as PR manipulation and bypassing code reviews (Decan et al., 2022; Saroar & Nayebi, 2023; Benedetti et al., 2022). GitHub CI combines CI workflows with the GitHub environment, generating issues related to privileges, permissions, and secrets (Koishybayev et al., 2022; Hilton et al., 2017; Benedetti et al., 2022).

**Existing DevSecOps Practices.** Security issues related to encryption, image signing, and vulnerability scanning remain in open-source DevSecOps environments (Kumar and Goyal, 2020). The SolarWinds incident showed that practices need more standard recommendations (Nadgowda and Luan, 2021; Williams, 2022). This can lead to incomplete toolsets and compromised software designs.

**Low-code Platforms.** Integrating low-code platforms such as PowerApps, AppSheet, and KiSSFLOW in DevOps may introduce security issues (Rafi et al., 2022).

**Software Supply Chain (SSC).** The unified design of the CI server in a CD pipeline poses security challenges, as attackers can compromise the entire system by altering one part (Throner et al., 2021; Bass et al., 2015; Ullah et al., 2017; Hilton et al., 2017). Automated SSCs can propagate human errors, such as not updating vulnerable dependencies, leading to pipeline breaks, for example, the Log4j attack (Enck and Williams, 2022; Byrne et al., 2020; Williams, 2022). Securing the build process is crucial since tools such as Tekton, Jenkins, GHA, Travis CI, and AWS Code Deploy are widely used (Enck and Williams, 2022; Karl et al., 2022). Failure to promptly update and address risks can result in intrusions, such as the SolarWinds attacks (Nadgowda and Luan, 2021; Williams, 2022).

## 5 ANALYSIS AND DISCUSSIONS

The provided list (RQ1) encompasses a diverse range of tools and technologies to enhance the security posture of CI/CD pipelines, primarily focusing on Docker-based cloud environments. This includes security scanning tools, automated testing frameworks, monitoring solutions, and vulnerability assessment and remediation tools, contributing to a robust and secure software development lifecycle.

Integrating these tools and technologies within CI/CD pipelines significantly enhances security by addressing vulnerabilities, ensuring code quality, and proactively monitoring and responding to security threats. For instance, tools such as Docker Bench for Security and SonarQube help identify and rectify security issues early in development. Meanwhile, monitoring tools such as Prometheus and Nagios provide real-time insights into the deployed applications' operational status and security posture.

The proposed (RQ2) tools and practices aim to bolster CI/CD pipeline security. Tools cover code analysis, dependency scanning, and runtime protection, while practices emphasise collaboration, automated testing, and secure software supply chains. Implementing these measures may enhance security, streamline processes, and mitigate risks in CI/CD pipelines; however, accurate tests are needed on cloud platforms.

The excerpt (RQ3) provides a comprehensive overview of the security challenges inherent in cloud-based CI/CD pipelines, summarised below: -
- Installation and updating issues,
- Practitioners and developers' issues,
- Organisational issues,
- Difficulties with third-party and OSS tools.

# 6 THREATS TO VALIDITY

In our systematic literature review (SLR), we identified potential threats to validity across several areas, including search strategy, data collection, study selection, and synthesis. We conducted automated searches using diverse terminology to accommodate various taxonomies, though some digital libraries were excluded due to complex search strings and irrelevant results. Our study selection process adhered to established guidelines from Zhang et al. (2011), Kitchenham et al. (2022), Brereton et al. (2007), and Wohlin (2014).

Based on Runeson and Höst's (2009) framework, we identified the following threats:

Internal Validity: Potential data extraction errors were mitigated by thorough double-checking.

External Validity: Strict criteria may have led to a higher exclusion rate, potentially introducing selection bias, but they were essential for relevance. Comprehensive search techniques helped minimise the risk of missing significant studies.

Construct Validity: Standardization efforts addressed inconsistencies in study definitions.

Reliability: Variability in study design and quality was a concern, though we aimed to include a diverse range of studies to reduce the impact of publication bias.

# 7 CONCLUSIONS AND FUTURE WORK

Our systematic literature review provided valuable insights into the existing methods, tools, and technologies (RQ1) for maintaining security in the CI/CD pipeline over the cloud platforms.

To keep up with the continually updating environment, practitioners and researchers should stay updated on the latest advancements where future research is needed.

We have uncovered various tools, frameworks, and practices (RQ2) proposed by researchers to fortify security in the CI/CD pipeline. With cloud platforms ubiquitous, these findings suggest significant insights for practitioners and future researchers aiming to stay at the cutting edge of secure DevOps practices.

Finally, we have reported the challenges and issues that arise when dealing with security considerations in cloud-based CI/CD pipelines (RQ3). These issues involved container vulnerability, lack of integration between security tools and IDEs, and dependency on third-party software and OSS tools. Close cooperation between practitioners, security specialists, and researchers is needed to mitigate the research gaps.

We aim to apply Topic Modeling (an unsupervised ML technique (Sefara and Rangata, 2023) that uses Natural Language Processing) methods such as Latent Semantic Analysis (LSA), Probabilistic Latent Semantic Analysis (pLSA), and Latent Dirichlet Allocation (LDA) effectively applied to analyse scattered and fragmented security-related text data (for example, plain text, lack of integration, disorganised contents, lack of contexts such as partial incident reports, truncated logs, or isolated pieces of information, etc. which can be derived from grey literature, and the industries).

We also aim to propose a blockchain-based solution (Akbar et al., 2022; Bankar and Shah, 2020) (an advanced database mechanism for maintaining data privacy) for addressing the insufficient container security (for example, beyond 80% of Docker hub images contain one high level of vulnerability discovered by researchers after scanning 300,000 images in 85,000 repositories) (Zhang et al., 2018; Shu et al., 2017), insecure deployment environments (such as updating vulnerable dependencies, a human error which leads to cyberattacks such as Log4j, SolarWinds, CodeCov etc.) (Benedetti et al., 2022b; Enck and Williams, 2022; Byrne et al., 2020; Karl et al., 2022), etc. Before this, we also aim to conduct a literature review on blockchain-based solutions for securing the CI/CD pipeline.

In conclusion, this SLR gave us an understanding of CI/CD security and plans for future works, combining methodologies and technologies to fortify the foundations of secure software integration and deployment in cloud platforms.

# REFERENCES

Á. Révész, and N. Pataki, "Containerized A/B Testing," Proc. of the Sixth Workshop on Software Quality Analysis, Monitoring, Improvement, and Applications (Belgrade, Serbia, September 11-13, 2017) SQAMIA'17, 2017, pp. 14(1)-14(8).

Abhishek, M. K., & Rao, D. R. (2021, July). Framework to secure docker containers. In 2021 Fifth World Conference on Smart Trends in Systems Security and Sustainability (WorldS4) (pp. 152-156). IEEE.

Akbar, M. A., Mahmood, S., & Siemon, D. (2022, June). Toward effective and efficient DevOps using blockchain. In Proceedings of the 26th International Conference on Evaluation and Assessment in Software Engineering (pp. 421-427).

Alfadel, M., Nagy, N. A., Costa, D. E., Abdalkareem, R., & Shihab, E. (2023). Empirical analysis of security-related code reviews in npm packages. Journal of Systems and Software, 203, 111752.

Alonso, J., Piliszek, R., & Cankar, M. (2022). Embracing IaC through the DevSecOps philosophy: Concepts, challenges, and a reference framework. IEEE Software, 40(1), 56-62.

Athamnah, M., Hussain, M. F., & Hasan, S. S. (2021, November). Impact of Running Dynamic/Static Scans on the Performance of an App Running in a GKE Clusters. In 2021 Second International Conference on Intelligent Data Science Technologies and Applications (IDSTA) (pp. 46-53). IEEE.

Bankar, S., & Shah, D. (2020, November). DevOps project artifacts management using blockchain technology. In ECAI&ML international conference (pp. 115-120).

Bar, P., Benfredj, R., Marks, J., Ulevinov, D., Wozniak, B., Casale, G., & Knottenbelt, W. J. (2013, April). Towards a monitoring feedback loop for cloud applications. In Proceedings of the 2013 international workshop on Multi-cloud applications and federated clouds (pp. 43-44).

Bass, L., Holz, R., Rimba, P., Tran, A. B., & Zhu, L. (2015, May). Securing a deployment pipeline. In 2015 IEEE/ACM 3rd International Workshop on Release Engineering (pp. 4-7). IEEE.

Benedetti, G., Verderame, L., & Merlo, A. (2022, November). Automatic security assessment of github actions workflows. In Proceedings of the 2022 ACM Workshop on Software Supply Chain Offensive Research and Ecosystem Defenses (pp. 37-45).

Benedetti, G., Verderame, L., & Merlo, A. (2022, September). Alice in (software supply) chains: risk identification and evaluation. In International Conference on the Quality of Information and Communications Technology (pp. 281-295). Cham: Springer International Publishing.

Bennett, B. T., & Barrett, M. L. (2018). Incorporating devops into undergraduate software engineering courses: A suggested framework. Journal of Computing Sciences in Colleges, 34(2), 180-187.

Berkovich, S., Kam, J., & Wurster, G. (2020). {UBCIS}: Ultimate benchmark for container image scanning. In 13th USENIX Workshop on Cyber Security Experimentation and Test (CSET 20).

Billawa, P., Bambhore Tukaram, A., Díaz Ferreyra, N. E., Steghöfer, J. P., Scandariato, R., & Simhandl, G. (2022, August). Sok: Security of microservice applications: A practitioners' perspective on challenges and best practices. In Proceedings of the 17th International Conference on Availability, Reliability and Security (pp. 1-10).

Boeker, M., Vach, W., & Motschall, E. (2013). Google Scholar as replacement for systematic literature searches: good relative recall and precision are not enough. BMC Medical Research Methodology, 13, 1-12.

Brady, K., Moon, S., Nguyen, T., & Coffman, J. (2020, January). Docker container security in cloud computing. In 2020 10th Annual Computing and Communication Workshop and Conference (CCWC) (pp. 0975-0980). IEEE.

Brereton, P., Kitchenham, B. A., Budgen, D., Turner, M., & Khalil, M. (2007). Lessons from applying the systematic literature review process within the software engineering domain. Journal of systems and software, 80(4), 571-583.

Byrne, A., Nadgowda, S., & Coskun, A. K. (2020, December). Ace: Just-in-time serverless software component discovery through approximate concrete execution. In Proceedings of the 2020 Sixth International Workshop on Serverless Computing (pp. 37-42).

Cankar, M., Petrovic, N., Pita Costa, J., Cernivec, A., Antic, J., Martincic, T., & Stepec, D. (2023, April). Security in DevSecOps: Applying Tools and Machine Learning to Verification and Monitoring Steps. In Companion of the 2023 ACM/SPEC International Conference on Performance Engineering (pp. 201-205).

Chen, L., Babar, M. A., & Zhang, H. (2010, April). Towards an evidence-based understanding of electronic data sources. At the 14th International Conference on Evaluation and Assessment in Software Engineering (EASE), BCS Learning & Development.

Chhillar, D., & Sharma, K. (2019, February). ACT Testbot and 4S Quality Metrics in XAAS Framework. In 2019 International Conference on Machine Learning, Big Data, Cloud and Parallel Computing (COMITCon) (pp. 503-509). IEEE.

Christakis, M., Cottenier, T., Filieri, A., Luo, L., Mansur, M. N., Pike, L., ... & Visser, W. (2022, November). Input splitting for cloud-based static application security testing platforms. In Proceedings of the 30th ACM Joint European Software Engineering Conference and Symposium on the Foundations of Software Engineering (pp. 1367-1378).

Davis, J. C., Amusuo, P., & Bushagour, J. R. (2022, May). A first offering of software engineering. In Proceedings of the First International Workshop on Designing and Running Project-Based Courses in Software Engineering Education (pp. 5-9).

Decan, A., Mens, T., Mazrae, P. R., & Golzadeh, M. (2022, October). On the use of github actions in software development repositories. In 2022 IEEE International Conference on Software Maintenance and Evolution (ICSME) (pp. 235-245). IEEE.

Drees, J. P., Gupta, P., Hüllermeier, E., Jager, T., Konze, A., Priesterjahn, C., ... & Somorovsky, J. (2021, November). Automated detection of side channels in cryptographic protocols: DROWN the ROBOTs!. In Proceedings of the 14th ACM Workshop on Artificial Intelligence and Security (pp. 169-180).

Düllmann, T. F., Paule, C., & van Hoorn, A. (2018, May). Exploiting devops practices for dependable and secure continuous delivery pipelines. In Proceedings of the 4th International Workshop on Rapid Continuous Software Engineering (pp. 27-30).

Dursun, H. (2023, June). Full Spec Software via Platform Engineering: Transition from Bolting-on to Building-in. In Proceedings of the 27th International Conference on Evaluation and Assessment in Software Engineering (pp. 172-175).

El Khairi, A., Caselli, M., Knierim, C., Peter, A., & Continella, A. (2022, November). Contextualizing system calls in containers for anomaly-based intrusion detection. In Proceedings of the 2022 on Cloud Computing Security Workshop (pp. 9-21).

Enck, W., & Williams, L. (2022). Top five challenges in software supply chain security: Observations from 30 industry and government organizations. IEEE Security & Privacy, 20(2), 96-100.

Faustino, J., Adriano, D., Amaro, R., Pereira, R., & da Silva, M. M. (2022). DevOps benefits: A systematic literature review. Software: Practice and Experience, 52(9), 1905-1926.

Fehlmann, T., & Kranich, E. (2021). ART for Agile: Autonomous Real-Time Testing in the Product Development Cycle. In Systems, Software and Services Process Improvement: 28th European Conference, EuroSPI 2021, Krems, Austria, September 1–3, 2021, Proceedings 28 (pp. 377-390). Springer International Publishing.

Fitzgerald, B., & Stol, K. J. (2014, June). Continuous software engineering and beyond: trends and challenges. In Proceedings of the 1st International Workshop on rapid continuous software engineering (pp. 1-9).

Fitzgerald, B., & Stol, K. J. (2017). Continuous software engineering: A roadmap and agenda. Journal of Systems and Software, 123, 176-189.

Flittner, M., Balaban, S., & Bless, R. (2016, April). Cloudinspector: A transparency-as-a-service solution for legal issues in cloud computing. In 2016 IEEE International Conference on Cloud Engineering Workshop (IC2EW) (pp. 94-99). IEEE.

Garg, S., & Garg, S. (2019, March). Automated cloud infrastructure, continuous integration and continuous delivery using docker with robust container security. In 2019 IEEE Conference on Multimedia Information Processing and Retrieval (MIPR) (pp. 467-470). IEEE.

Gruhn, V., Hannebauer, C., & John, C. (2013, August). Security of public continuous integration services. In Proceedings of the 9th International Symposium on open collaboration (pp. 1-10).

Gusenbauer, M., & Haddaway, N. R. (2020). Which academic search systems are suitable for systematic reviews or meta‐analyses? Evaluating retrieval qualities of Google Scholar, PubMed, and 26 other resources. Research synthesis methods, 11(2), 181-217.

Hilton, M., Nelson, N., Tunnell, T., Marinov, D., & Dig, D. (2017, August). Trade-offs in continuous integration: assurance, security, and flexibility. In Proceedings of the 2017 11th Joint Meeting on Foundations of Software Engineering (pp. 197-207).

Huang, M., Fan, W., Huang, W., Cheng, Y., & Xiao, H. (2020, June). Research on building exploitable vulnerability database for cloud-native app. In 2020 IEEE 4th Information Technology, Networking, Electronic and Automation Control Conference (ITNEC) (Vol. 1, pp. 758-762). IEEE.

Hudic, A., Flittner, M., Lorünser, T., Radl, P. M., & Bless, R. (2016, August). Towards a unified secure cloud service development and deployment life-cycle. In 2016 11th International Conference on Availability, Reliability and Security (ARES) (pp. 428-436). IEEE.

Humble, J., & Farley, D. (2010). Continuous delivery: reliable software releases through build, test, and deployment automation. Pearson Education..

Jamshidi, P., Pahl, C., Mendonça, N. C., Lewis, J., & Tilkov, S. (2018). Microservices: The journey so far and challenges ahead. IEEE Software, 35(3), 24-35.

Kang, H., Le, M., & Tao, S. (2016, April). Container and microservice driven design for cloud infrastructure devops. In 2016 IEEE International Conference on Cloud Engineering (IC2E) (pp. 202-211). IEEE.

Karl, M., Musch, M., Ma, G., Johns, M., & Lekies, S. (2022, October). No keys to the kingdom required: a comprehensive investigation of missing authentication vulnerabilities in the wild. In Proceedings of the 22nd ACM Internet Measurement Conference (pp. 619-632).

Kitchenham, B. (2004). Procedures for performing systematic reviews. Keele, UK, Keele University, 33(2004), 1-26.

Kitchenham, B. (2006). Evidence-based software engineering and systematic literature reviews. In Product-Focused Software Process Improvement: 7th International Conference, PROFES 2006, Amsterdam, The Netherlands, June 12-14, 2006. Proceedings 7 (pp. 3-3). Springer Berlin Heidelberg.

Kitchenham, B. A., Dyba, T., & Jorgensen, M. (2004, May). Evidence-based software engineering. In Proceedings. 26th International Conference on Software Engineering (pp. 273-281). IEEE.

Kitchenham, B., Madeyski, L., & Budgen, D. (2022). How should software engineering secondary studies include grey material?. IEEE Transactions on Software Engineering, 49(2), 872-882.

Kitchenham, B., Madeyski, L., & Budgen, D. (2022). SEGRESS: Software engineering guidelines for reporting secondary studies. IEEE Transactions on Software Engineering, 49(3), 1273-1298.

Koishybayev, I., Nahapetyan, A., Zachariah, R., Muralee, S., Reaves, B., Kapravelos, A., & Machiry, A. (2022). Characterizing the security of github {CI} workflows. In 31st USENIX Security Symposium (USENIX Security 22) (pp. 2747-2763).

Kumar, R., & Goyal, R. (2020). Modeling continuous security: A conceptual model for automated DevSecOps using open-source software over cloud (ADOC). Computers & Security, 97, 101967.

Lacoste, F. J. (2009, August). Killing the gatekeeper: Introducing a continuous integration system. In 2009 agile conference (pp. 387-392). IEEE.

Le, M. V., Ahmed, S., Williams, D., & Jamjoom, H. (2023, July). Securing container-based clouds with syscall-aware scheduling. In Proceedings of the 2023 ACM Asia Conference on Computer and Communications Security (pp. 812-826).

Leite, L., Rocha, C., Kon, F., Milojicic, D., & Meirelles, P. (2019). A survey of DevOps concepts and challenges. ACM Computing Surveys (CSUR), 52(6), 1-35.

Leppänen, M., Mäkinen, S., Pagels, M., Eloranta, V. P., Itkonen, J., Mäntylä, M. V., & Männistö, T. (2015). The highways and country roads to continuous deployment. Ieee software, 32(2), 64-72.

Li, Z., & Rainer, A. (2022, November). Academic search engines: constraints, bugs, and recommendations. In Proceedings of the 13th International Workshop on Automating Test Case Design, Selection and Evaluation (pp. 25-32).

Lopes, N., Martins, R., Correia, M. E., Serrano, S., & Nunes, F. (2020, December). Container hardening through automated seccomp profiling. In Proceedings of the 2020 6th International Workshop on Container Technologies and Container Clouds (pp. 31-36).

Luo, L., Schäf, M., Sanchez, D., & Bodden, E. (2021, August). Ide support for cloud-based static analyses. In Proceedings of the 29th ACM Joint meeting on european software engineering conference and symposium on the foundations of software engineering (pp. 1178-1189).

Mahboob, J., & Coffman, J. (2021, January). A kubernetes ci/cd pipeline with asylo as a trusted execution environment abstraction framework. In 2021 IEEE 11th Annual Computing and Communication Workshop and Conference (CCWC) (pp. 0529-0535). IEEE.

Maplesden, D., Tempero, E., Hosking, J., & Grundy, J. C. (2015). Performance analysis for object-oriented software: A systematic mapping. IEEE Transactions on Software Engineering, 41(7), 691-710.

Nadgowda, S., & Luan, L. (2021, December). tapiserí: Blueprint to modernize DevSecOps for real world. In Proceedings of the Seventh International Workshop on Container Technologies and Container Clouds (pp. 13-18).

Newkirk, J. (2002, May). Introduction to agile processes and extreme programming. In Proceedings of the 24th international conference on Software engineering (pp. 695-696).

Ohm, M., Sykosch, A., & Meier, M. (2020, August). Towards detection of software supply chain attacks by forensic artifacts. In Proceedings of the 15th international conference on availability, reliability and security (pp. 1-6).

Okafor, C., Schorlemmer, T. R., Torres-Arias, S., & Davis, J. C. (2022, November). Sok: Analysis of software supply chain security by establishing secure design properties. In Proceedings of the 2022 ACM Workshop on Software Supply Chain Offensive Research and Ecosystem Defenses (pp. 15-24).

Pan, Z., Shen, W., Wang, X., Yang, Y., Chang, R., Liu, Y., ... & Ren, K. (2023). Ambush From All Sides: Understanding Security Threats in Open-Source Software CI/CD Pipelines. IEEE Transactions on Dependable and Secure Computing, 21(1), 403-418.

Pashchenko, I., Scandariato, R., Sabetta, A., & Massacci, F. (2021, May). Secure software development in the era of fluid multi-party open software and services. In 2021 IEEE/ACM 43rd International Conference on Software Engineering: New Ideas and Emerging Results (ICSE-NIER) (pp. 91-95). IEEE.

Pecka, N., Ben Othmane, L., & Valani, A. (2022, May). Privilege escalation attack scenarios on the devops pipeline within a kubernetes environment. In Proceedings of the International Conference on Software and System Processes and International Conference on Global Software Engineering (pp. 45-49).

Petticrew, M., & Roberts, H. (2008). Systematic reviews in the social sciences: A practical guide. John Wiley & Sons.

Rafi, S., Akbar, M. A., Sánchez-Gordón, M., & Colomo-Palacios, R. (2022, September). Devops practitioners' perceptions of the low-code trend. In Proceedings of the 16th ACM/IEEE International Symposium on Empirical Software Engineering and Measurement (pp. 301-306).

Rajapakse, R. N., Zahedi, M., Babar, M. A., & Shen, H. (2022). Challenges and solutions when adopting DevSecOps: A systematic review. Information and software technology, 141, 106700.

Révész, Á., & Pataki, N. (2019, March). Continuous A/B testing in containers. In Proceedings of the 2019 2nd International Conference on Geoinformatics and Data Analysis (pp. 11-14).

Romero, E. E., Camacho, C. D., Montenegro, C. E., Acosta, Ó. E., Crespo, R. G., Gaona, E. E., & Martínez, M. H. (2022). Integration of DevOps practices on a noise monitor system with CircleCI and Terraform. ACM Transactions on Management Information Systems (TMIS), 13(4), 1-24.

Runeson, P., & Höst, M. (2009). Guidelines for conducting and reporting case study research in software engineering. Empirical software engineering, 14, 131-164.

Saboor, A., Hassan, M. F., Akbar, R., Susanto, E., Shah, S. N. M., Siddiqui, M. A., & Magsi, S. A. (2022). Root-Of-Trust for Continuous Integration and Continuous Deployment Pipeline in Cloud Computing. Computers, Materials and Continua, 73(2), 2223-2239.

Saroar, S. G., & Nayebi, M. (2023, June). Developers' perception of GitHub Actions: A survey analysis. In Proceedings of the 27th International Conference on Evaluation and Assessment in Software Engineering (pp. 121-130).

Sefara, T. J., & Rangata, M. R. (2023, August). Topic classification of tweets in the broadcasting domain using machine learning methods. In 2023 International Conference on Artificial Intelligence, Big Data, Computing and Data Communication Systems (icABCD) (pp. 1-6). IEEE.

Shahin, M., Babar, M. A., & Zhu, L. (2017). Continuous integration, delivery and deployment: a systematic review on approaches, tools, challenges and practices. IEEE access, 5, 3909-3943.

Shahin, M., Babar, M. A., Zahedi, M., & Zhu, L. (2017, November). Beyond continuous delivery: an empirical investigation of continuous deployment challenges. In 2017 ACM/IEEE International Symposium on Empirical Software Engineering and Measurement (ESEM) (pp. 111-120). IEEE.

Shahin, M., Rezaei Nasab, A., & Ali Babar, M. (2023). A qualitative study of architectural design issues in DevOps. Journal of Software: Evolution and Process, 35(5), e2379.

Shahin, M., Rezaei Nasab, A., & Ali Babar, M. (2023). A qualitative study of architectural design issues in DevOps. Journal of Software: Evolution and Process, 35(5), e2379.

Shahin, M., Zahedi, M., Babar, M. A., & Zhu, L. (2019). An empirical study of architecting for continuous delivery and deployment. Empirical Software Engineering, 24, 1061-1108.

Shu, R., Gu, X., & Enck, W. (2017, March). A study of security vulnerabilities on docker hub. In Proceedings of the Seventh ACM on Conference on Data and Application Security and Privacy (pp. 269-280).

Sokolowski, D., Weisenburger, P., & Salvaneschi, G. (2021, August). Automating serverless deployments for DevOps organizations. In Proceedings of the 29th ACM Joint Meeting on European Software Engineering Conference and Symposium on the Foundations of Software Engineering (pp. 57-69).

Ståhl, D., & Bosch, J. (2014). Modeling continuous integration practice differences in industry software development. Journal of Systems and Software, 87, 48-59..

Staples, M., & Niazi, M. (2007). Experiences using systematic review guidelines. Journal of Systems and Software, 80(9), 1425-1437.

Throner, S., Hütter, H., Sänger, N., Schneider, M., Hanselmann, S., Petrovic, P., & Abeck, S. (2021, August). An advanced devops environment for microservice-based applications. In 2021 IEEE International Conference on Service-Oriented System Engineering (SOSE) (pp. 134-143). IEEE.

Torkura, K. A., Sukmana, M. I., & Meinel, C. (2017, December). Integrating continuous security assessments in microservices and cloud native applications. In Proceedings of the10th International Conference on Utility and Cloud Computing (pp. 171-180).

Tu, W., Wei, Y. H., Antichi, G., & Pfaff, B. (2021, August). Revisiting the open vswitch dataplane ten years later. In Proceedings of the 2021 ACM SIGCOMM 2021 Conference (pp. 245-257).

Ullah, F., Raft, A. J., Shahin, M., Zahedi, M., & Babar, M. A. (2017). Security support in continuous deployment pipeline. arXiv preprint arXiv:1703.04277.

Waseem, M., Liang, P., Ahmad, A., Khan, A. A., Shahin, M., Abrahamsson, P., ... & Mikkonen, T. (2023). Understanding the Issues, Their Causes and Solutions in Microservices Systems: An Empirical Study. arXiv preprint arXiv:2302.01894.

Waseem, M., Liang, P., Shahin, M., Ahmad, A., & Nassab, A. R. (2021, June). On the nature of issues in five open source microservices systems: An empirical study. In Proceedings of the 25th International Conference on Evaluation and Assessment in Software Engineering (pp. 201-210).

Weber, I., Nepal, S., & Zhu, L. (2016). Developing dependable and secure cloud applications. IEEE Internet Computing, 20(3), 74-79.

Wohlin, C. (2014, May). Guidelines for snowballing in systematic literature studies and a replication in software engineering. In Proceedings of the 18th International Conference on Evaluation and Assessment in Software Engineering (pp. 1-10).

Zampetti, F., Nardone, V., & Di Penta, M. (2022, May). Problems and solutions in applying continuous integration and delivery to 20 open-source cyber-physical systems. In Proceedings of the 19th International Conference on Mining Software Repositories (pp. 646-657).

Zampetti, F., Tamburri, D., Panichella, S., Panichella, A., Canfora, G., & Di Penta, M. (2023). Continuous integration and delivery practices for cyber-physical systems: An interview-based study. ACM Transactions on Software Engineering and Methodology, 32(3), 1-44.

Zhang, H., Babar, M. A., & Tell, P. (2011). Identifying relevant studies in software engineering. Information and Software Technology, 53(6), 625-637.

Zhang, Y., Vasilescu, B., Wang, H., & Filkov, V. (2018, October). One size does not fit all: an empirical study of containerized continuous deployment workflows. In Proceedings of the 2018 26th ACM Joint Meeting on European Software Engineering Conference and Symposium on the Foundations of Software Engineering (pp. 295-306).

Zhu, C., Zhang, M., Wu, X., Xu, X., & Li, Y. (2023). Client-specific upgrade compatibility checking via knowledge-guided discovery. ACM Transactions on Software Engineering and Methodology, 32(4), 1-31.